# DNA nanotechnology meets nanophotonics

Na Liu

2nd Physics Institute, University of Stuttgart, Pfaffenwaldring 57, 70569 Stuttgart, Germany
Max Planck Institute for Solid State Research, Heisenbergstrasse 1, 70569 Stuttgart, Germany
Email: na.liu@pi2.uni-stuttgart.de

**Call-out sentence:** It will be very constructive, if more research funds become available to support young researchers with bold ideas and meanwhile allow for failures and contingent outcomes.

The first time I heard the two terms 'DNA nanotechnology' and 'nanophotonics' mentioned together was from Paul Alivisatos, who delivered the Max Planck Lecture in Stuttgart, Germany, on a hot summer day in 2008. In his lecture, Paul showed how a plasmon ruler containing two metallic nanoparticles linked by a DNA strand could be used to monitor nanoscale distance changes and even the kinetics of single DNA hybridization events in real time, readily correlating nanoscale motion with optical feedback.[1] Until this day, I still vividly remember my astonishment by the power and beauty of these two nanosciences, when rigorously combined together.

In the past decades, DNA has been intensely studied and exploited in different research areas of nanoscience and nanotechnology. At first glance, DNA-based nanophotonics seems to deviate quite far from the original goal of Nadrian Seeman, the founder of DNA nanotechnology, who hoped to organize biological entities using DNA in high-resolution crystals. As a matter of fact, DNA-based nanophotonics does closely follow his central spirit. That is, apart from being a genetic material for inheritance, DNA is also an ideal material for building molecular devices.

A great leap forward along the direction of DNA-based nanophotonics emerged, attributed to the revolutionary invention of DNA origami by Paul Rothemund in 2006.[2] The formation of DNA origami involves the folding of a long scaffold by hundreds of short staple strands into arbitrary 2D and 3D shapes. DNA origami offers much higher rigidity than discrete DNA strands. Most importantly, it allows for the organization of individual functional groups on a single template with unprecedented precision, sequence specificity, addressability, and programmability.[3] Thereafter, DNA nanotechnology entered a flourishing period, in which functionalizations of different nanoscale elements on DNA origami, ranging from organic to inorganic materials were successively realized. In particular, the essential building blocks for nanophotonics, such as metallic nanocrystals, fluorophores, quantum dots, upconversion nanoparticles, among others were accurately assembled on DNA origami with high yields. Noteworthily, the successful assembly of anisotropic metallic nanocrystals on origami, such as gold nanorods by Hao Yan's group in 2011,[4] paved the road towards the realization of complex nanophotonic architectures with tailored optical functionalities.

When approaching the 15th anniversary of DNA origami, many crucial milestones have been accomplished for nanophotonics. To name a few, DNA origami has enabled the fabrication of truly 3D nanophotonic architectures operating at visible frequencies.[5] This remains challenging for other state of the art nanofabrication approaches, especially for top-down nanotechniques. Furthermore, DNA origami has empowered reconfigurable nanophotonic structures, whose optical responses can be modified by a variety of external stimuli, including pH, temperature, light, deoxyribozymes, among others.[6] Meanwhile, this has led to advanced optical sensors for detection of miscellaneous molecular species, including proteins, RNA, ATP, cocaine, etc. These examples take inherent advantage of DNA, which is sensitive to a broad range of chemical modifications. Along this line, DNA origami has served as unique platform to push the sensing limit towards the single molecule level. For instance, the precise and quantitative positioning of individual molecules inside the hotspots of plasmonic nanostructures on DNA origami can give rise to highly reproducible surface-enhanced Raman spectroscopy signals.[7] Also, DNA origami has helped to quantify the interplay between single emitters and plasmonic nanostructures, providing great insights into the underlying physics of light-matter interactions on the nanoscale.[8] Apparently, there are too many remarkable examples to elucidate in this short viewpoint. Rather, I would like to end this paragraph by highlighting the recent groundbreaking works by Paul Rothemund's group, who has combined the respective strengths of both DNA nanotechnology and lithographic patterning to map nanocavity emission via precision placement of DNA origami[9] as well as to optimize nanophotonic device performance via control over the fluorescent dipole orientations using DNA origami.[10]

So what do we aim for 20 years along the road? For fundamental research, smart strategies need to be developed for creation of large DNA origami templates to accommodate multiple functional elements or photonic objects with substantial sizes. In particular, we need to devote more endeavors to developing functionalization protocols for the assembly of large photonic objects with strong optical responses (e.g., gold nanorods of 100-200 nm in length) on DNA origami with high yield and high fidelity, because this will bring about profound significance to eliminate many current constraints in designing DNA-based nanophotonic devices. Very likely, artificial intelligence will be employed not only for molecular programming and computational designs, but also for fully automated synthesis and assembly.

For applied research, comprehensive investigations on robust molecular adaptors on DNA origami and rapid control schemes for reconfiguration of DNA origami will be very instructive to foster a new class of active nanophotonic devices. In the long-term perspective, complex nanophotonic architectures with advanced functionalities could be envisioned. Integrated photonic circuits, such as optical wireless links comprising transmitter and receiver antennas together with localized single emitters could be entirely templated by DNA origami and operate in a fully dynamic fashion. Assembly lines or even artificial nanofactories could be built to produce molecular compounds in resemblance to the working principles of photosynthesis systems, taking advantage of both the biochemical activities of DNA and efficient energy transport mechanisms offered by nanophotonic units. For therapeutic and clinical applications, more efficient tumour regression schemes in vivo could be developed by encapsulation of both therapeutic molecules and plasmonic nanoparticles inside DNA origami carriers for robotic drug delivery in combination with photothermal therapy.[11,12]

Equally important, how should we further promote the development of this research field? Breakthrough findings in any field rely on the people, who develop and execute the ideas. In my viewpoint, it will be very constructive, if more research funds become available to support young researchers with bold ideas and meanwhile allow for failures and contingent outcomes. These ideas might contain counterintuitive hypotheses, unconventional methodologies and risky approaches. As we know, in the history of a science many groundbreaking concepts were often not accepted at the beginning. One should always remain open-minded.

**ORCID**

Na Liu: 0000-0001-5831-3382

**Notes**

The author declares no competing financial interest.

**Acknowledgements**

This project was supported by the European Research Council (ERC *Dynamic Nano*) grant and the Max Planck Society (Max Planck Fellow).